\documentclass[11pt,twocolumn]{article}

\usepackage{amsmath}
\usepackage{amssymb}
\usepackage{graphicx}
\usepackage{booktabs}
\usepackage{hyperref}
\usepackage{geometry}
\title{\textbf{A Comparative Study of Dynamic Programming and Reinforcement Learning in Finite-Horizon Dynamic Pricing}}

\author{Lev Razumovskiy \and Nikolay Karenin}
\newcommand{\Addresses}{{
\bigskip
\footnotesize

Lev Razumovskiy	, \textsc{RAMAX Group}\par\nopagebreak
  \textit{E-mail address} : \text{lev.razumovskiy@ramax.com}

\medskip

Nikolay Karenin, \textsc{RAMAX Group}\par\nopagebreak
  \textit{E-mail address} : \text{nikolay.karenin@ramax.com}

}}
\date{}

\begin{document}

\twocolumn[
\maketitle
\begin{abstract}
This paper provides a systematic comparison between Fitted Dynamic Programming (DP), where demand is estimated from data, and Reinforcement Learning (RL) methods in finite-horizon dynamic pricing problems. 
We analyze their performance across environments of increasing structural complexity, ranging from a single typology benchmark to multi-typology settings with heterogeneous demand and inter-temporal revenue constraints. 
Unlike simplified comparisons that restrict DP to low-dimensional settings, we apply dynamic programming in richer, multi-dimensional environments with multiple product types and constraints.
We evaluate revenue performance, stability, constraint satisfaction behavior, and computational scaling, highlighting the trade-offs between explicit expectation-based optimization and trajectory-based learning.
\end{abstract}
\vspace{0.5cm}
]

\section{Introduction}

\subsection{Motivation}

Dynamic pricing is a central problem in real estate, airline revenue management, e-commerce, and inventory control. Applications also include ticket sales for live events, where pricing decisions must balance stochastic demand and limited seat capacity over a fixed selling horizon.

The objective is to determine an optimal pricing strategy over a finite horizon in order to maximize expected revenue under stochastic demand and inventory constraints. Classical finite-horizon formulations are rooted in dynamic programming (DP), where optimal controls are computed from Bellman recursions under explicit demand models \cite{gallego1994optimal}. This line of work is systematized in revenue management references \cite{talluri2004theory, phillips2005pricing}.

At the same time, practical pricing systems rarely have a perfectly known demand model. This motivates \emph{learning-while-optimizing} approaches, where prices are chosen both to earn revenue and to reduce demand uncertainty. Foundational regret-based analyses quantify this exploration--exploitation trade-off and provide near-optimal learning guarantees \cite{besbes2009dynamic, farias2010prior}. Data-driven implementations that combine demand estimation and optimization further demonstrate strong practical relevance in retail settings \cite{ferreira2018analytics}.

A major challenge is scalability. Even when demand learning is incorporated, exact DP becomes computationally difficult as the number of products, state variables, or operational constraints grows. This limitation is emphasized both in survey work and in algorithmic research on controlled exploration and high-dimensional pricing \cite{denboer2015dynamic, denboer2014controlled, javanmard2019highdim}.

In parallel, reinforcement learning (RL) has become a promising alternative because it can learn directly from trajectory data without explicit enumeration of transition probabilities. Early model-free pricing studies in non-stationary environments and more recent competition-oriented settings indicate that RL can handle market dynamics that are hard to encode analytically \cite{rana2014realtime, kastius2022competition}. Recent finite-horizon comparisons between RL and data-driven DP confirm this potential, but also show that relative performance depends strongly on problem structure and data budget~\cite{lange2025reinforcement}.

Compared with Lange et al.~\cite{lange2025reinforcement}, we consider a different type of problem complexity. Their study focuses on monopoly and duopoly airline pricing with customer-choice effects and competition, but still centers on a single typology setting. In contrast, our environments introduce multiple typologies that are optimized jointly under shared revenue constraints, resulting in a genuinely coupled multi-dimensional control problem. Another important difference concerns the Fitted DP benchmark itself. In Lange et al.~\cite{lange2025reinforcement}, the data-driven DP procedure estimates sales from data and then converts these estimates into probabilities via a Poisson approximation, even though the underlying purchase process is modeled using a Binomial distribution. In our setting, Fitted DP is more closely aligned with the underlying demand specification, which makes the comparison with RL methodologically cleaner.

Despite substantial progress in dynamic pricing, there is still limited systematic evidence on how model-based DP and model-free RL compare in finite-horizon settings as state-action dimensionality increases. To close this gap, we design a sequence of environments with increasing structural complexity, from a single typology benchmark to multi-typology settings. In multi-typology environments, we additionally introduce cumulative revenue constraints (implemented via penalty terms) to create more operationally realistic and non-trivial decision regimes. This allows us to study not only aggregate revenue, but also how each method adapts its policy structure under tighter inter-temporal requirements.

\subsection{Aim of the Study}

Despite extensive literature on DP and RL in dynamic pricing, there is limited systematic comparison across environments with increasing structural complexity.

We address the following questions:

\begin{itemize}
    \item How does DP performance scale as dimensionality increases?
    \item When does RL become computationally preferable?
    \item How do global constraints alter the relative behavior of DP and RL?
\end{itemize}

\subsection{Our Contribution}

 Our contribution is empirical and methodological rather than algorithmic:

  \begin{itemize}
      \item We construct a unified benchmark with four finite-horizon environments with varied state-action dimensionality and constrained multi-typology
  settings.
      \item We perform a like-for-like comparison of Fitted DP and model-free RL under aligned action discretization, demand assumptions, and training budgets.
      \item Beyond final revenue, we analyze stability and constraint-related behavior (including revenue distributions before the penalty step), which reveals differences in how methods adapt under inter-temporal requirements.
  \end{itemize}

\section{Problem Formulation}

\subsection{Typologies and Notation}

All environments considered in this paper consist of one or multiple typologies. 
Each typology represents an independent inventory-demand pair. 
In more complex environments, typologies are coupled through shared constraints such as cumulative revenue targets.

We consider $K$ typologies indexed by $i=1,\dots,K$ over a finite horizon $t=1,\dots,T$.

For each typology we denote:

\begin{itemize}
    \item $p_{i, t}$ — chosen price,
    \item $Q_{i, t}$ — stochastic demand realization,
    \item $N_{i, t}$ — remaining inventory.
\end{itemize}

When only a single typology is considered we will omit index $i$ for convenience.

Sales are inventory constrained:

\[
S_{i, t} = \min(Q_{i, t}, N_{i, t-1}), \quad N_{i, t} = N_{i, t-1} - S_{i, t}.
\]

Here we denoted the initial inventory by $N_i=N_{i,0}$. 

\subsection{Revenue and Penalty}

Immediate revenue is computed as:

\[
r_t^{\text{rev}} = \sum_{i=1}^K p_{i, t} S_{i, t}.
\]

If a cumulative revenue target $R^{\text{target}}$ must be satisfied at time $\tau$, a penalty is applied:
\[
\text{penalty}_\tau = 
\mu \max(0, R^{\text{target}} - R_\tau),
\]
where

\[
R_t = \sum_{\ell=1}^{t} r_\ell^{\text{rev}}
\]
and $\mu$ is the penalty severity coefficient. In this case, the total reward at the constraint time is computed as:

\[
r_\tau = r_\tau^{\text{rev}} - \text{penalty}_\tau.
\]

If there is no revenue constraint at time $t$, we simply set $r_t=r_t^{\text{rev}}$.

For $K>1$, the penalty depends on aggregate revenue across all typologies, so the immediate reward is non-separable and decisions are coupled across typologies.

\subsection{Demand Model}

At each time period $t = 1,\dots,T$ and for each typology $i$, demand is modeled as a random variable

\[
Q_{i, t} \sim \text{Poisson}\big(d_i(p_{i, t},t)\big),
\]
where $d_i(p_{i, t},t)$ is the demand intensity (arrival rate).

\paragraph{Interpretation.}

The Poisson specification implies:

\begin{itemize}
    \item Demand arrives as a counting process.
    \item $\mathbb{E}[Q_{i, t}] = \mathrm{Var}(Q_{i, t}) = d_i(p_{i, t},t)$.
    \item Conditional on the intensity, arrivals within a period are independent.
\end{itemize}

This formulation is standard in revenue management and dynamic pricing, 
where customer arrivals during a short time interval are well approximated 
by a Poisson process.

\subsubsection*{Demand Intensity Function}

In experiments, demand intensity depends on both price and time:

\[
d_i(p,t)
=
\max\big(0, \alpha_i - \beta_i p\big)
\left(1 + \frac{t}{T}\right).
\]

The structure consists of two components:

\begin{enumerate}
    \item \textbf{Price sensitivity term:}
    \[
    \alpha_i - \beta_i p,
    \]
    where
    \begin{itemize}
        \item $\alpha_i > 0$ represents the baseline demand level,
        \item $\beta_i > 0$ controls price elasticity.
    \end{itemize}
    
    In other words, demand decreases linearly with price. The truncation
\[
\max(0, \alpha_i - \beta_i p)
\]
ensures non-negativity of the Poisson rate.
    
    \item \textbf{Time scaling term:}
    \[
    1 + \frac{t}{T},
    \]
    which introduces deterministic growth over time.
    As $t$ increases, demand intensity rises linearly,
    reflecting increasing urgency to buy.
\end{enumerate}

Demands across typologies are assumed independent.

\subsection{MDP Formulation}
We now reformulate the problem within the reinforcement learning framework.

The states and actions are given by the following table:

\begin{center}
\begin{tabular}{lcc}
\toprule
 & State & Action \\
\midrule
Env 1 & $(t,N_{1,t})$ &  $(p_t)$\\
Env 2 & $(t,N_{1,t},N_{2,t}, R_t)$ & $(p_{1,t},p_{2,t})$\\
Env 3 & $(t,N_{1,t},N_{2,t}, R_t)$ & $(p_{1,t},p_{2,t})$\\
Env 4 & $(t,N_{1,t}, R_t)$ & $(p_t)$\\
\bottomrule
\end{tabular}
\end{center}
Most of it is defined by the number of typologies with one exception -- cumulative revenue $R_t$ is not needed in the first environment since there are no revenue constraints in it. 

Transition formulas were already discussed before. Reward coincides with step revenue $r_t$ and discount-factor $\lambda=1$ in all environments. This gives us cumulative revenue maximization problem 
$$R_T\to \max$$

Since DP requires a finite action space, we use discretized price grids, i.e. 
\[
p_{1, t} \in \mathcal{P}_i,\quad |\mathcal{P}_i|<\infty.
\]
DP also requires the number of states to be finite. This leads to the necessary discretization of cumulative revenue in DP. The discretization step is chosen to be 0.1 for environment 4 and 0.5 for environments 2 and 3 due to computational constraints.

\section{Solution Methods}

\subsection{Dynamic Programming with Demand Estimation}

\subsubsection{Demand Estimation}

Demand is estimated via linear regression on data collected from random episodes, using realized sales as the target variable:

\begin{equation}
d(p, t) = \beta_0 + \beta_1 \cdot p + \beta_2 \cdot \left(\frac{t}{T}\right) 
+ \beta_3\cdot  p \cdot \left(\frac{t}{T}\right).
\end{equation}

Observations from sold-out periods are excluded from demand estimation. 

\subsubsection{Bellman recursion}
The DP algorithm computes the optimal value function

\[
V_t^*(s), 
\]
where $s$ is a state. We start from states with $t=T$ and set $V_{T}^*(s)=0$. Then we go back to $t=0$ and compute $V_t^*(s)$ using Bellman recursion:

\[
V_t^*(s)
=
\max_{a\in A}
\sum_{s'}
\mathbb{P}(s \to s'| a)
\left[
r_t
+
V_{t+1}^*(s')
\right], 
\]
where we denoted a set of all possible actions by $A$.

\subsubsection{Computational Complexity}

The computational complexity of DP is primarily determined by the size of the discretized state--action space. Let $|S|$ denote the typical number of feasible states per time step and let $|A|$ denote the number of admissible actions. For each state--action pair, the Bellman update requires summation over all feasible successor states. In the worst case, their number is of order $|S|$, which yields an additional multiplicative factor $|S|$ on top of the $T\cdot |S|\cdot |A|$ state--action evaluations. Therefore, the overall complexity is of order
\[
\mathcal{O}(T\cdot |S|^2\cdot |A|).
\]

In the unconstrained single typology case, the state is $(t,N_{1,t})$, so the number of states grows linearly with the number of inventory levels. In constrained multi-typology environments, the state additionally includes cumulative revenue, and the state-space size grows multiplicatively with the inventory levels of all typologies and with the number of revenue bins. For instance, with two typologies and discretized cumulative revenue,
\[
|S| \approx N_1\cdot N_2\cdot |R|,
\]
where $|R|$ is the number of admissible discretized revenue levels. Thus, the computational burden of DP in our setting is driven mainly by the rapid growth of the discretized multi-dimensional state space rather than by the horizon length alone.

Note that the actual training time remains almost unchanged as the number of training episodes increases.

\subsection{Reinforcement Learning}

\subsubsection{Models and Training}
We apply model-free RL algorithms (DQN, PPO, A2C) using vectorized environments. All algorithms use the default hyperparameters provided by Stable-Baselines3.

The RL algorithm set was chosen to match the discrete action formulation required by DP. These algorithms represent complementary families for discrete control: value-based (DQN) and policy-gradient/actor-critic (A2C, PPO). Our conclusions are scoped to this representative subset rather than all RL methods.

Training is performed in episodes. 
Each algorithm interacts with four parallel environments.
The model observes state transitions and updates policy/value networks via stochastic gradient optimization.

For DQN in two-typology environments, the joint discrete action $(p_{1,t}, p_{2,t})$ is encoded as a single categorical action.

Unlike DP, RL:

\begin{itemize}
    \item does not enumerate demand realizations,
    \item does not compute explicit expectations,
    \item approximates value functions from sampled trajectories.
\end{itemize}

\subsubsection{Computational Complexity}

Unlike DP, RL does not admit a closed-form complexity in terms of the state and action space sizes, as it relies on sampling rather than exhaustive enumeration. Instead, its computational cost is primarily driven by the number of training episodes, the episode horizon, and the cost of policy evaluation and update steps.

More formally, the overall complexity can be expressed as
\[
  \mathcal{O}\bigl(N_{\text{episodes}} \cdot T \cdot C_{\text{model}}\bigr),
\]
where $N_{\text{episodes}}$ is the number of training episodes, $T$ is the episode length and $C_{\text{model}}$ denotes the per-episode cost of forward and backward passes through all neural networks this model uses.

A detailed analysis of $C_{\text{model}}$, which depends on the architecture and input dimensionality, is beyond the scope of this paper.

In practice, the number of episodes required for convergence may itself depend on the problem structure, which makes RL complexity difficult to characterize analytically.

\subsection{Environment 1: Single Typology}

This environment serves as a baseline setting with a single product typology and no constraints.
        

The action is a discrete price:
\[
p \in \mathcal{P}=\{0.5, 0.6, \dots, 2.0\}, \quad |\mathcal{P}| = 16.
\]


Demand follows a Poisson distribution with intensity
\[
d(p,t)
=
\max(0, 2 - p)
\left(1 + \frac{t}{T}\right).
\]

Starting inventory is $N=10$.
\subsection{Environment 2: Two Identical Typologies with Constraint}

This environment extends the baseline to two typologies that are identical to the one in Environment 1, but introduces a revenue constraint.

The action is a pair of prices
\[
a_t = (p_1, p_2),
\]
where each $p_i$ belongs to the same grid as in Environment 1.  
The total number of actions is $16^2 = 256$.


Each typology follows the same demand model as in Environment 1:
\[
d_i(p,t)
=
\max(0, 2 - p)
\left(1 + \frac{t}{T}\right), \quad i=1,2.
\]


A penalty with $R^{target}=12$ and $\mu=4$ is applied after $7$ steps:
\[
\text{penalty}_7 = 4\max(0, 12 - R_7).
\]

The initial inventories satisfy $N_1=N_2=6$.
\subsection{Environment 3: Two Heterogeneous Typologies}

This environment generalizes Environment 2 by introducing heterogeneity across typologies.

Prices are selected independently for each typology but from different ranges:
\[
p_1 \in [0.5, 1.6], \quad
p_2 \in [1.0, 3.0].
\]

The joint action space again contains $256$ actions.


Demand functions differ across typologies:
\[
d_1(p,t)
=
\max(0, 2.5 - 1.5p)
\left(1 + \frac{t}{T}\right),
\]
\[
d_2(p,t)
=
\max(0, 3.0 - 0.8p)
\left(1 + \frac{t}{T}\right).
\]

These differences reflect heterogeneity in price sensitivity and baseline demand.
A penalty with $R^{target}=12$ and $\mu=4$ is applied after $7$ steps as in environment 2:
\[
\text{penalty}_7 = 4\max(0, 12 - R_7).
\]
Note that the constraint is less restrictive here.

The initial inventories satisfy $N_1=N_2=6$.

\subsection{Environment 4: Single Typology with Constraint}

This environment isolates the effect of the constraint in the simplest setting.
Action space,  demand and starting inventory are the same as in Environment 1. But a penalty $R^{target}=5.5$ and $\mu=1$ is applied after $5$ steps:
\[
\text{penalty}_5= \max(0, 5.5 - R_5).
\]

\subsection{Environment Comparison}

\begin{center}
\begin{tabular}{lcccc}
\toprule
 & Env 1 & Env 2 & Env 3 & Env 4\\
\midrule
Typologies & 1 & 2 & 2 & 1\\
Price symmetry & -- & Yes & No & -- \\
Revenue penalty & No & Yes & Yes & Yes \\
Action dimension & 1 & 2 & 2 & 1\\
State dimension & 2 & 4 & 4 & 3\\
\bottomrule
\end{tabular}
\end{center}

\subsection{Result robustness}
To further assess the robustness of the results, we additionally consider an environment with a non-linear demand dependency on time and price. The detailed description and results for this setting are provided in Appendix~\ref{app:env5}.

\section{Results}

Each model was trained independently on each environment for different numbers of training episodes 
$$40, 100, 200, 400, 1000, 2000$$
with $10$ independent runs per setting.
Models were trained independently for each number of training episodes.  

The expected revenue for trained models was computed directly in environment $1$ and using $10000$ simulations in other environments. Plots show average acquired values and their standard deviations over 10 different training runs.   

Our main conclusions are further supported by experiments with non-linear demand, reported in Appendix~\ref{app:env5}.

\subsection{Results: Environment 1 (Single Typology)}

\begin{figure}[h]
    \centering
    \includegraphics[width=\columnwidth]{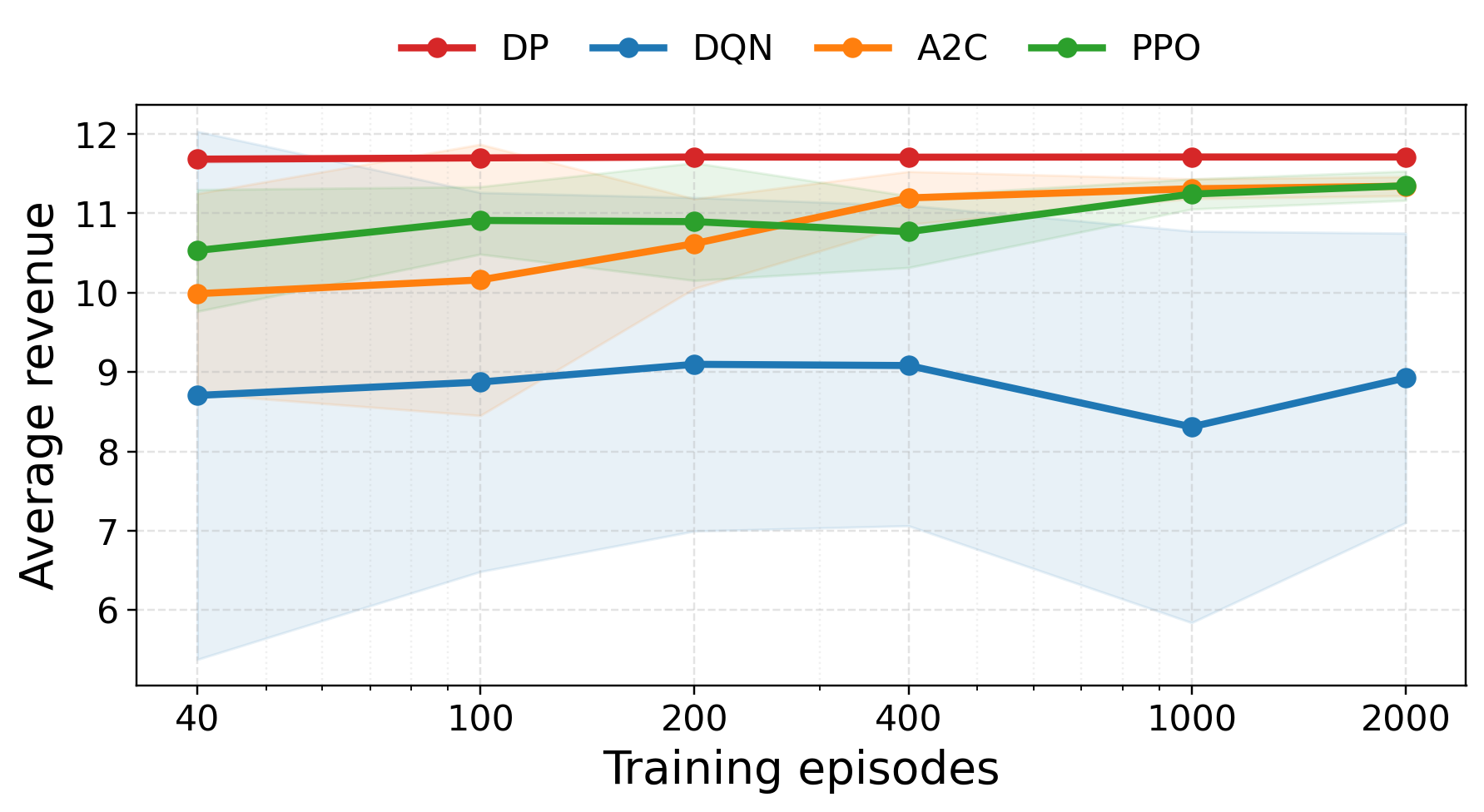}
    \caption{Environment 1}
    \label{fig:1_std}
\end{figure}

Figure~\ref{fig:1_std} presents the mean revenue together with one standard deviation bands for DQN, A2C, PPO, and Fitted DP as a function of the number of training episodes (log-scale).

Across all training budgets, Fitted DP consistently achieves the highest and most stable performance. The variability band is narrow, reflecting the deterministic nature of the Fitted dynamic programming procedure and the availability of the model structure.

For small training budgets (40--100 episodes), all reinforcement learning algorithms significantly underperform relative to the DP benchmark. In this regime, the variance is substantial, particularly for DQN, which exhibits both lower mean revenue and a wide standard deviation band. PPO demonstrates comparatively strong performance and moderate variability at this early stage.

In the intermediate regime (200--400 episodes), A2C stabilizes around a revenue level close to the DP benchmark and shows the fastest convergence among the RL methods. PPO does not show noticeable improvements. DQN remains less stable, with noticeable fluctuations in performance.

For large training budgets (1000 and 2000 episodes), A2C and PPO converge toward the DP benchmark. DQN fails to reach comparable revenue levels and continues to exhibit high variability.

Overall, the results indicate that in the single typology environment, policy-gradient methods---particularly PPO---provide the best trade-off between convergence speed, stability, and final performance. Value-based DQN converges more slowly and demonstrates higher variance, while model-based Fitted DP retains a slight performance advantage.

\subsection{Results: Environment 2 (Two Identical Typologies)}

\begin{figure}[h]
    \centering
    \includegraphics[width=\columnwidth]{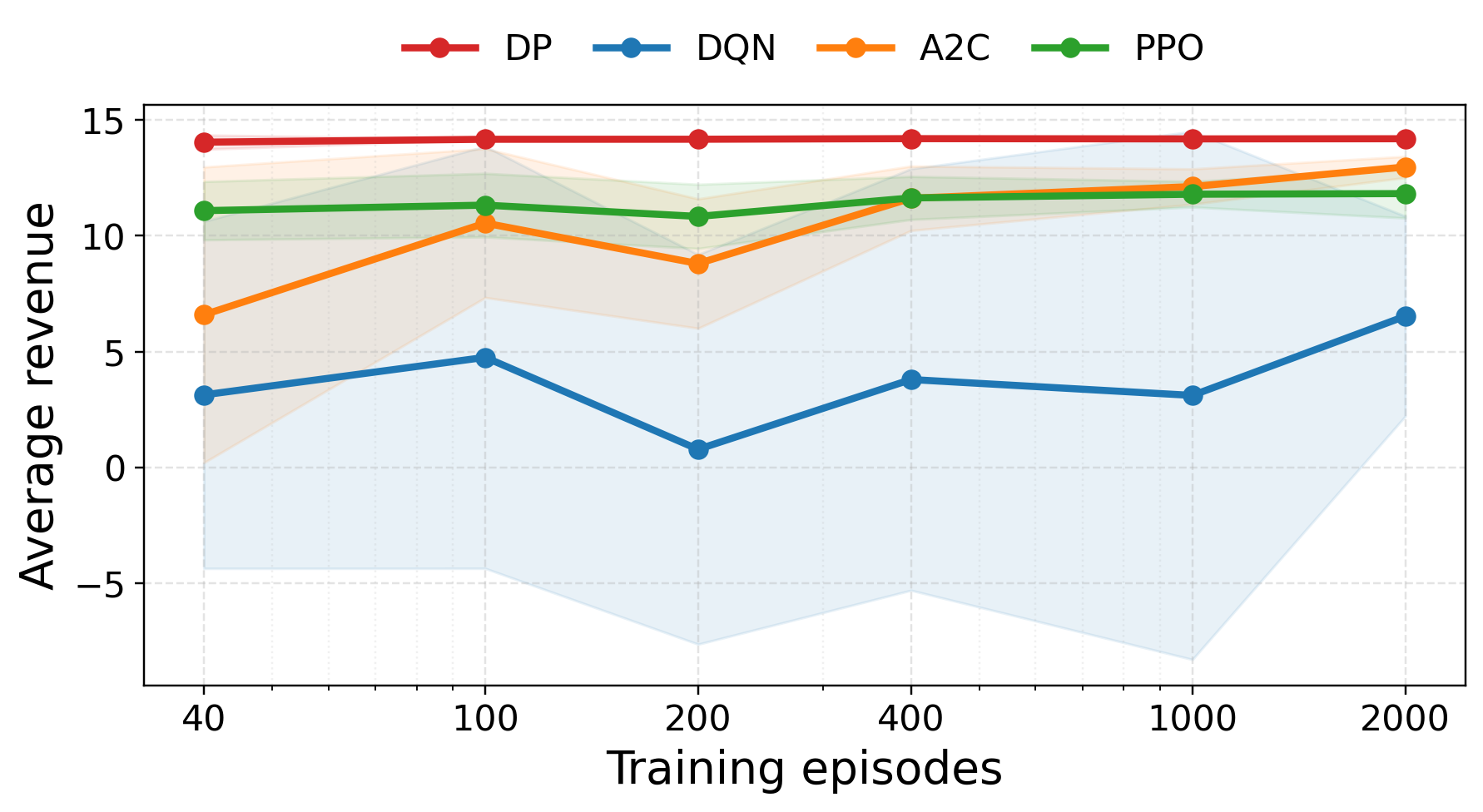}
    \caption{Environment 2: Two identical typologies}
    \label{fig:2_std}
\end{figure}

Figure~\ref{fig:2_std} presents the mean revenue together with one standard deviation bands for A2C, PPO, DQN, and Fitted DP in the environment with two identical typologies.

As in the single typology case, Fitted DP achieves the highest and most stable performance across all training budgets. The variability band remains narrow and almost constant, reflecting the deterministic optimization over the estimated demand model and the absence of sampling-based approximation error.

For small training budgets (40--100 episodes), A2C and DQN substantially underperform relative to the DP benchmark. A2C exhibits particularly high variance in this regime, with a wide standard deviation band indicating unstable early training. PPO demonstrates more stable behavior, achieving higher mean revenue and noticeably smaller variability even with limited data.

In the intermediate regime (200--400 episodes), PPO and A2C improve considerably. PPO stabilizes close to the DP benchmark and maintains relatively tight standard deviation bands. A2C shows significant improvement compared to the small-sample regime but still exhibits larger dispersion and slightly lower mean revenue than PPO. Meanwhile some DQN training runs achieve similar results, other training runs finish with far lower expected revenue. 

For larger training budgets (1000--2000 episodes), A2C almost reaches Fitted DP performance and shows low variability. PPO shows noticeable but small improvement. DQN also shows some improvement but fails to reach comparable revenue levels. Although neither RL method fully matches the deterministic DP performance, the gap becomes relatively small at higher training budgets.

Compared to Environment 1, the two-typology setting increases the dimensionality of both the action and state spaces, making the optimization problem more complex. This leads to slower stabilization for RL methods, especially in the low-data regime. Nevertheless, PPO and A2C continue to demonstrate strong scalability and robustness, suggesting that policy-gradient methods adapt well to moderate increases in problem dimensionality.

Overall, in the two identical typologies environment, Fitted DP retains a performance advantage due to explicit expectation computation, while PPO provides the best trade-off among the reinforcement learning methods in terms of convergence speed, stability, and final revenue.

\paragraph{Revenue distribution at the constraint step.}

\begin{figure}[h]
    \centering
    \includegraphics[width=\columnwidth]{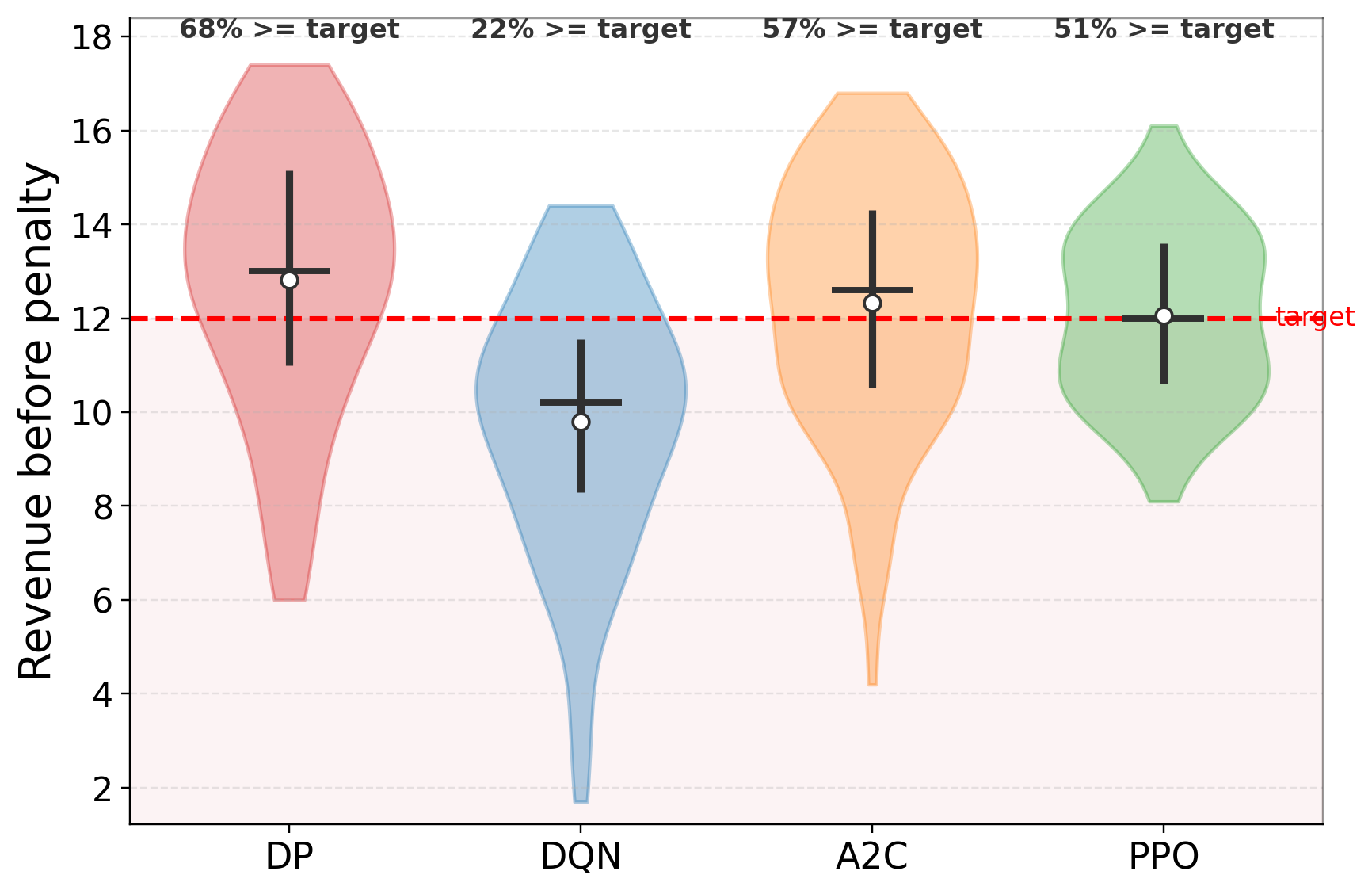}
    \caption{Distribution of cumulative revenue before the penalty step ($t=7$) for A2C, PPO, DQN, and Fitted DP in Environment 2. The dashed line denotes the revenue target.}
    \label{fig:penalty_violin_env2}
\end{figure}

To further analyze how different policies behave relative to the revenue constraint, we examine the distribution of cumulative revenue immediately before the penalty is applied (step $t=7$). Figure~\ref{fig:penalty_violin_env2} shows violin plots of the revenue distribution for A2C, PPO, DQN, and Fitted DP. The dashed horizontal line indicates the revenue target used in the constraint.

The distributions reveal noticeable differences in how the algorithms approach the constraint. 
The Fitted DP policy produces the highest median revenue before the penalty step and the largest 
concentration of probability mass above the revenue target, indicating that it satisfies the 
constraint with high probability.

A2C shows visually similar performance, but satisfies the constraint $11\%$ less often. 

Among the reinforcement learning methods, PPO generates revenues that are relatively close to the
constraint boundary. Its median lies approximately around the target level, and a noticeable
portion of the distribution still falls below the threshold. This indicates that while PPO often
approaches the required revenue level before the constraint step, it still triggers the penalty
almost half the time.

The DQN distribution is shifted further downward relative to PPO. Its median lies below the
target level, and a larger share of the probability mass is concentrated under the constraint
threshold. This suggests that DQN more frequently fails to reach the required revenue before
the penalty step and incurs the penalty most of the time.

Overall, while PPO and A2C exhibit larger dispersion than the DP solution, they tend to 
operate closer to the constraint boundary, whereas DQN more often underachieves the revenue 
target at the penalty step.

\subsection{Results: Environment 3 (Two Different Typologies)}

We now report the results of the numerical experiment in an environment with two distinct typologies, each characterized by its own demand function and price range. This setting introduces a higher-dimensional decision space with more intricate structure than environment $2$ and allows us to evaluate the robustness of different pricing approaches under heterogeneous demand conditions.

\begin{figure}[h]
    \centering
    \includegraphics[width=\columnwidth]{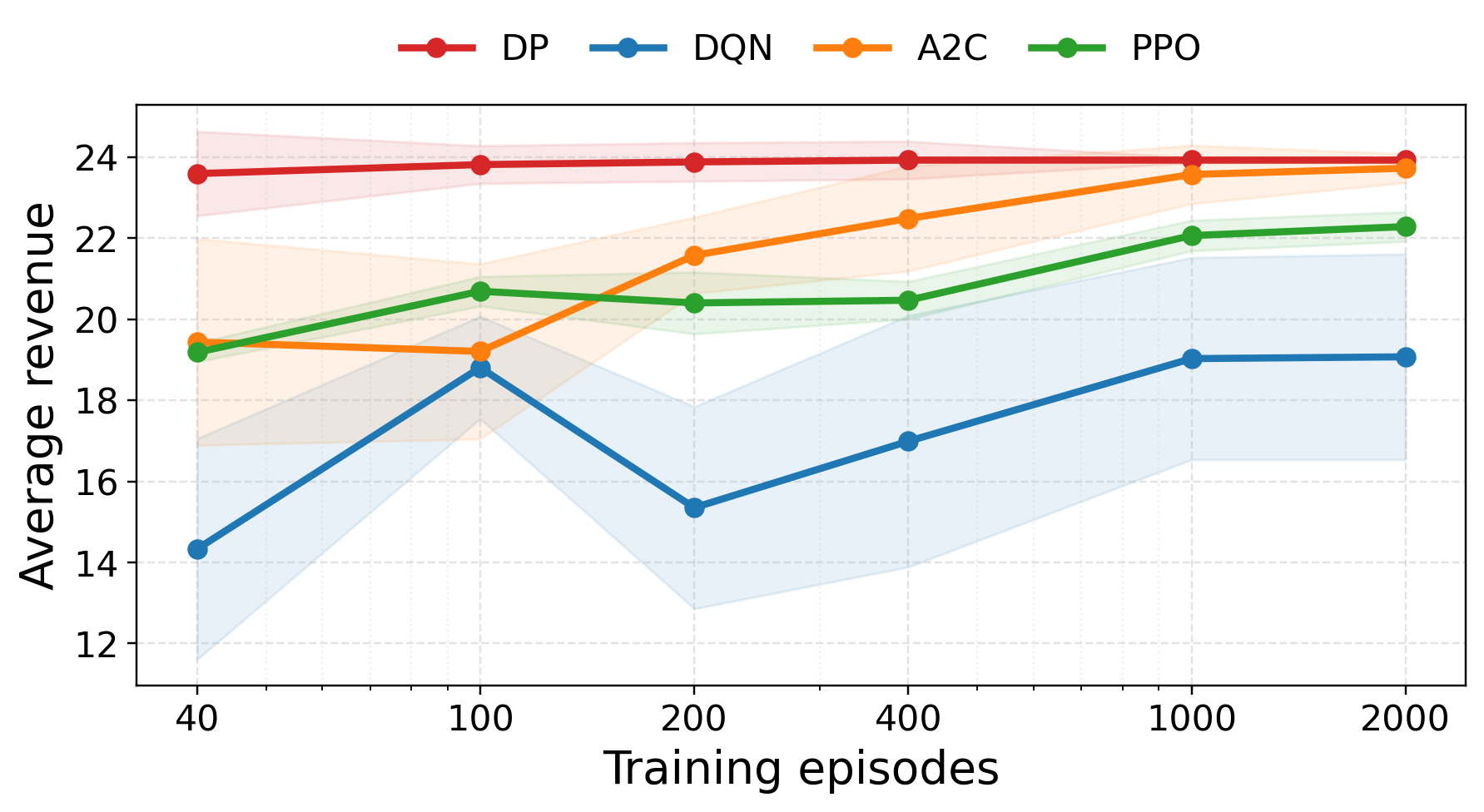}
    \caption{Environment 3: Two different typologies}
    \label{fig:3_std}
\end{figure}

Figure ~\ref{fig:3_std} presents the average revenue achieved by A2C, PPO, DQN, and the Fitted DP benchmark as a function of the number of training episodes. The shaded regions correspond to one standard deviation around the mean.

Several important observations emerge from the results.

First, the Fitted DP policy achieves strong performance even with limited data, providing a stable and relatively high baseline revenue. This is expected, as DP directly exploits the estimated demand model and optimizes decisions accordingly.

Second, reinforcement learning methods improve with the number of training episodes. In particular, A2C exhibits a steady increase in performance, seemingly reaching the DP benchmark. PPO shows a less stable learning trajectory, with lower performance at large training budgets but still significant improvement as the number of episodes increases. DQN again significantly underperforms all other methods. 

Third, at sufficiently large training budgets (around 2000 episodes), the RL methods become competitive with the Fitted DP approach. In some training runs, A2C slightly outperforms DP in terms of revenue. PPO, on the other hand, achieves lower performance but with more stable behavior.

Finally, the variance of A2C decreases as the number of training episodes increases, indicating improved policy stability. However, other RL methods show almost constant variance.

Overall, the results demonstrate that while Fitted DP provides a strong baseline in structured environments, reinforcement learning methods are capable of achieving comparable performance given sufficient training data, even in more complex multi-typology settings.

\paragraph{Revenue distribution at the constraint step (two typologies).}

\begin{figure}[h]
    \centering
    \includegraphics[width=\columnwidth]{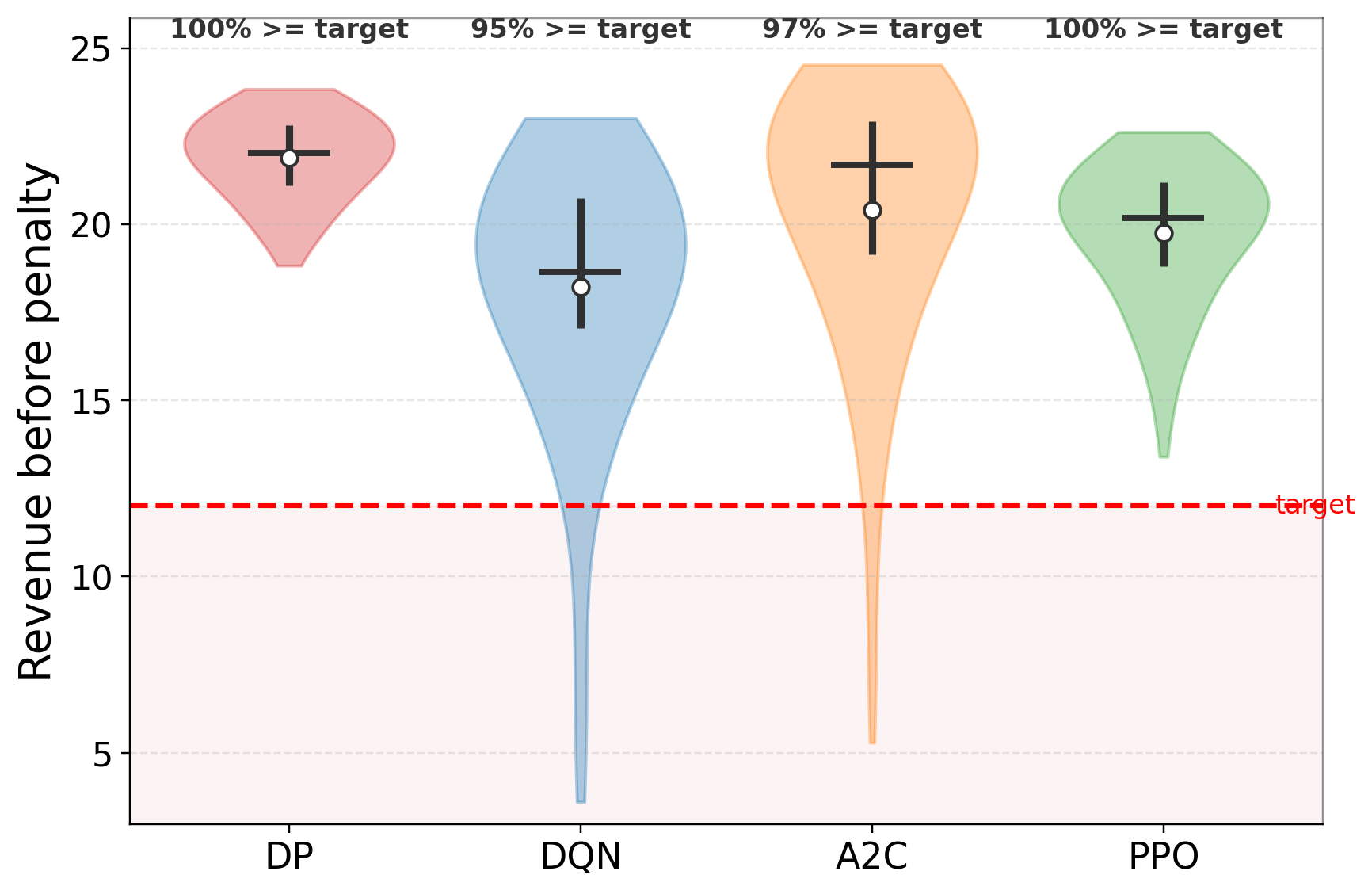}
    \caption{Distribution of cumulative revenue before the penalty step for A2C, PPO, DQN and Fitted DP in Environment 3 (two different typologies). The dashed line denotes the revenue target.}
    \label{fig:penalty_violin_env3}
\end{figure}

We now examine the distribution of cumulative revenue immediately before the penalty is applied in the environment with two different typologies. Figure~\ref{fig:penalty_violin_env3} presents violin plots for A2C, PPO, DQN, and Fitted DP. The dashed horizontal line indicates the revenue target.

The distributions highlight clear differences in how the methods handle the constraint in a higher-dimensional setting. The Fitted DP policy again shows the most concentrated distribution above the revenue target, indicating a high probability of satisfying the constraint.

Among the RL methods, A2C exhibits a wider spread with a noticeable lower tail, suggesting occasional significant underperformance relative to the average. PPO produces a much more concentrated distribution. And DQN show performance similar to A2c but with overall lower results.

Overall, all RL methods demonstrate increased variability compared to DP in this more complex environment. While PPO appears more stable, A2C shows greater dispersion and higher average revenue.

\subsection{Results: Environment 4 (Single-Typology Environment with Constraint)}

\begin{figure}[h]
    \centering
    \includegraphics[width=\columnwidth]{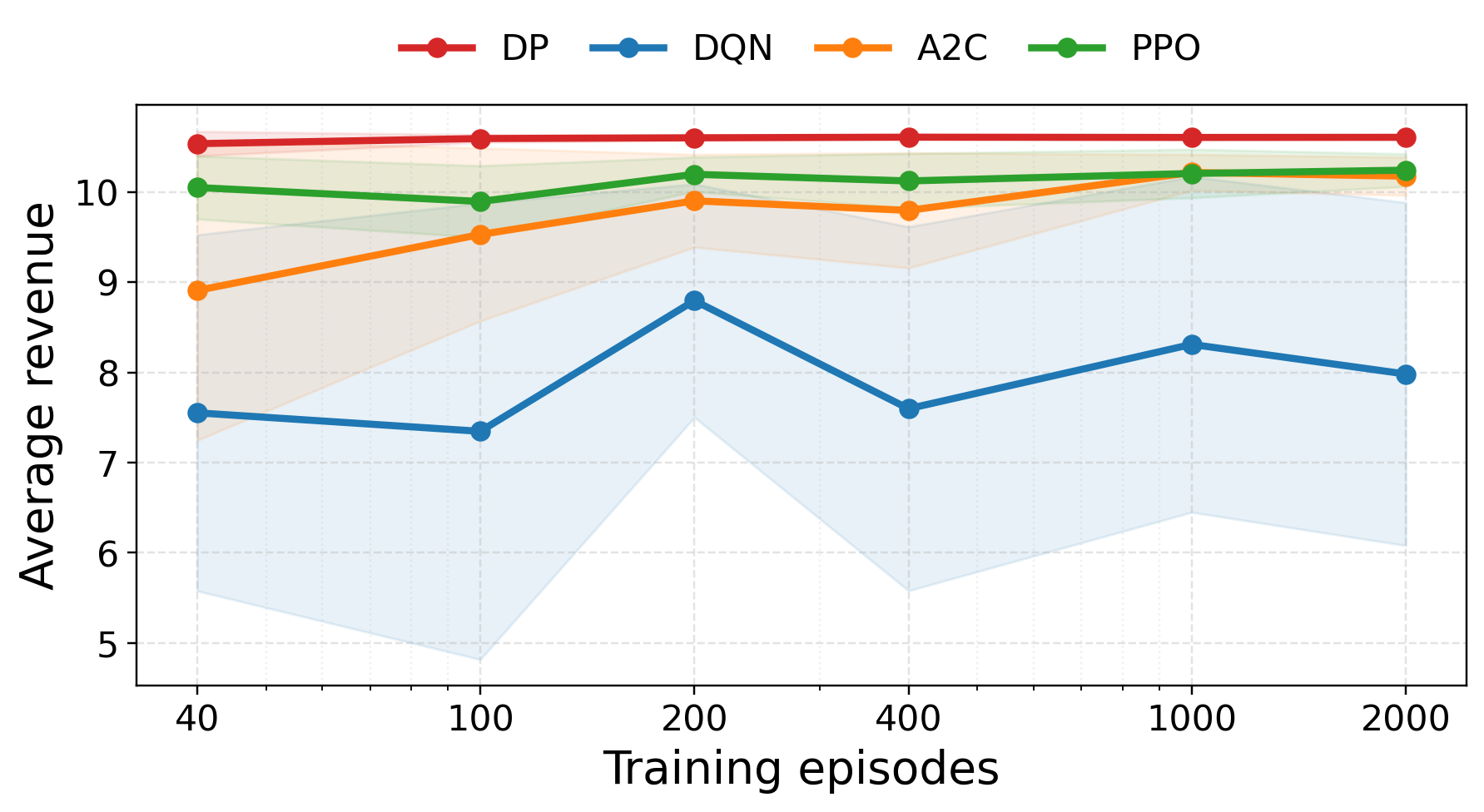}
    \caption{Environment 4: Single typology with constraint}
    \label{fig:4_std}
\end{figure}

Figure ~\ref{fig:4_std} shows the average revenue achieved by DQN, A2C, PPO, and the Fitted dynamic programming (DP) benchmark as a function of the number of training episodes. The shaded regions represent one standard deviation around the mean over multiple simulation runs.

Several key observations can be made.

First, the Fitted DP policy provides a stable and strong benchmark, consistently achieving the highest revenue across all training regimes. Its average performance remains almost constant.

Second, reinforcement learning methods exhibit different levels of stability and convergence behavior. PPO demonstrates the most consistent performance among RL methods, maintaining relatively stable revenue across different training budgets and gradually approaching the DP benchmark.

Third, A2C shows significant improvement as the number of training episodes increases. While its performance is initially lower, it becomes competitive with DP at higher training budgets, indicating its ability to learn constraint-aware pricing behavior over time.

In contrast, DQN exhibits high variance and unstable learning dynamics. Its performance fluctuates substantially across different training regimes, suggesting that value-based methods may struggle in environments with binding constraints and stochastic rewards.

Overall, the results indicate that while Fitted DP remains a reliable and robust benchmark in constrained environments, and policy-gradient methods such as PPO and A2C are capable of learning competitive strategies given sufficient training. 

\subsection{Training time}
Table \ref{tab:training_time_2000} reports the average training time in seconds for different algorithms across all environments.

\begin{table}[htbp]
\centering
\caption{Training time at $2000$ episodes (in seconds)}
\label{tab:training_time_2000}
\begin{tabular}{lrrrr}
\hline
 & DP & DQN & A2C & PPO \\
\hline
Env 1 & 0.2 & 4 & 6 & 16 \\
Env 2 & 1208 & 5 & 8 & 22 \\
Env 3 & 1211 & 5 & 8 & 22 \\
Env 4 & 178 & 4 & 6 & 16 \\
\hline
\end{tabular}
\end{table}

RL methods remain computationally efficient even at $2000$ training episodes. A moderate increase in training time is observed for environments $2$ and $3$, which have larger action spaces. Among RL methods, PPO consistently requires more training time than DQN and A2C.

Fitted DP training time is largely independent of the number of episodes, whereas RL methods scale approximately linearly with it. As a result, for small training budgets (e.g., $40$ episodes), DP exhibits comparable training time. However, in more complex environments, DP becomes at least an order of magnitude slower than RL methods even at $2000$ episodes.

\section{Conclusion}

In this paper, we studied a finite-horizon dynamic pricing problem under stochastic demand and inventory constraints, and conducted a systematic comparison between model-based dynamic programming (DP) and reinforcement learning (RL) approaches across several environments of increasing complexity.

The results consistently demonstrate that Fitted dynamic programming provides a strong and reliable benchmark. Due to its explicit expectation computation, DP achieves stable performance with low variance across all settings, including multi-typology environments and scenarios with revenue constraints.

At the same time, reinforcement learning methods show the ability to learn competitive pricing strategies given sufficient training data. Among the considered algorithms, PPO exhibits the most stable performance and gives the best results in low training budgets, while A2C demonstrates better results at larger training budgets. Meanwhile DQN shows both slower convergence and higher variance, especially in constrained environments.

The experiments highlight several important patterns. First, the gap between RL and DP is most pronounced in low-data regimes, where RL methods suffer from high variance and insufficient exploration. Second, as the number of training episodes increases, RL approaches gradually close this gap and can achieve near-optimal performance. Third, increasing the dimensionality of the problem---for example, by introducing multiple typologies---slows down learning and increases variability a bit, but does not prevent RL methods from eventually converging to strong policies.

In constrained environments, the learning problem becomes more challenging, as algorithms must simultaneously optimize revenue and satisfy the constraint. This leads to increased variability and highlights the importance of stability in policy learning. In such settings, PPO and A2C demonstrate the ability to learn constraint-aware strategies, while value-based methods such as DQN appear less robust.

Overall, our findings suggest that while Fitted DP remains a powerful tool when a reliable demand model is available, reinforcement learning provides a flexible and scalable alternative that can perform competitively in complex environments, particularly when sufficient training data is available.

Future research directions include extending the framework to other demand models, incorporating additional operational constraints, and studying hybrid approaches that combine model-based and data-driven methods.

\appendix
\section{Additional Environment with Non-linear Demand}
\label{app:env5}

To further assess the robustness of the comparison between Fitted DP and RL methods, we consider an additional environment with a non-linear demand dependency on time and price. The purpose of this experiment is to verify whether the main conclusions of the paper remain valid beyond the linear demand structures used in the main text.

\subsection{Environment 5: Single Typology with Non-linear Demand}

This environment is based on the single typology setting from environment 1, but replaces the linear demand specification with a non-linear one. The action space remains unchanged:
\[
p \in \mathcal{P}=\{0.5, 0.6, \dots, 2.0\}, \quad |\mathcal{P}| = 16.
\]

The starting inventory is again $N=10$, and the time horizon is $T=10$.

Demand follows a Poisson distribution with intensity
\[
d(p,t) = L(t)\cdot v(p),
\]
where 
\[
L(t) = 1.05\cdot \exp(0.01\cdot t^2 - 0.032\cdot t + 0.16),
\]
\[
v(p) = \max(0, 4.7\cdot exp(0.2(2-p)-1)).
\]

\subsection{Modifications}

To account for the non-linear demand specification, the demand estimation model used in Fitted DP is augmented with quadratic terms. Specifically, the features $t^2$ and $p^2$ are added to improve the quality of the approximation. 

\subsection{Results}

\begin{figure}[htbp]
    \centering
    \includegraphics[width=\linewidth]{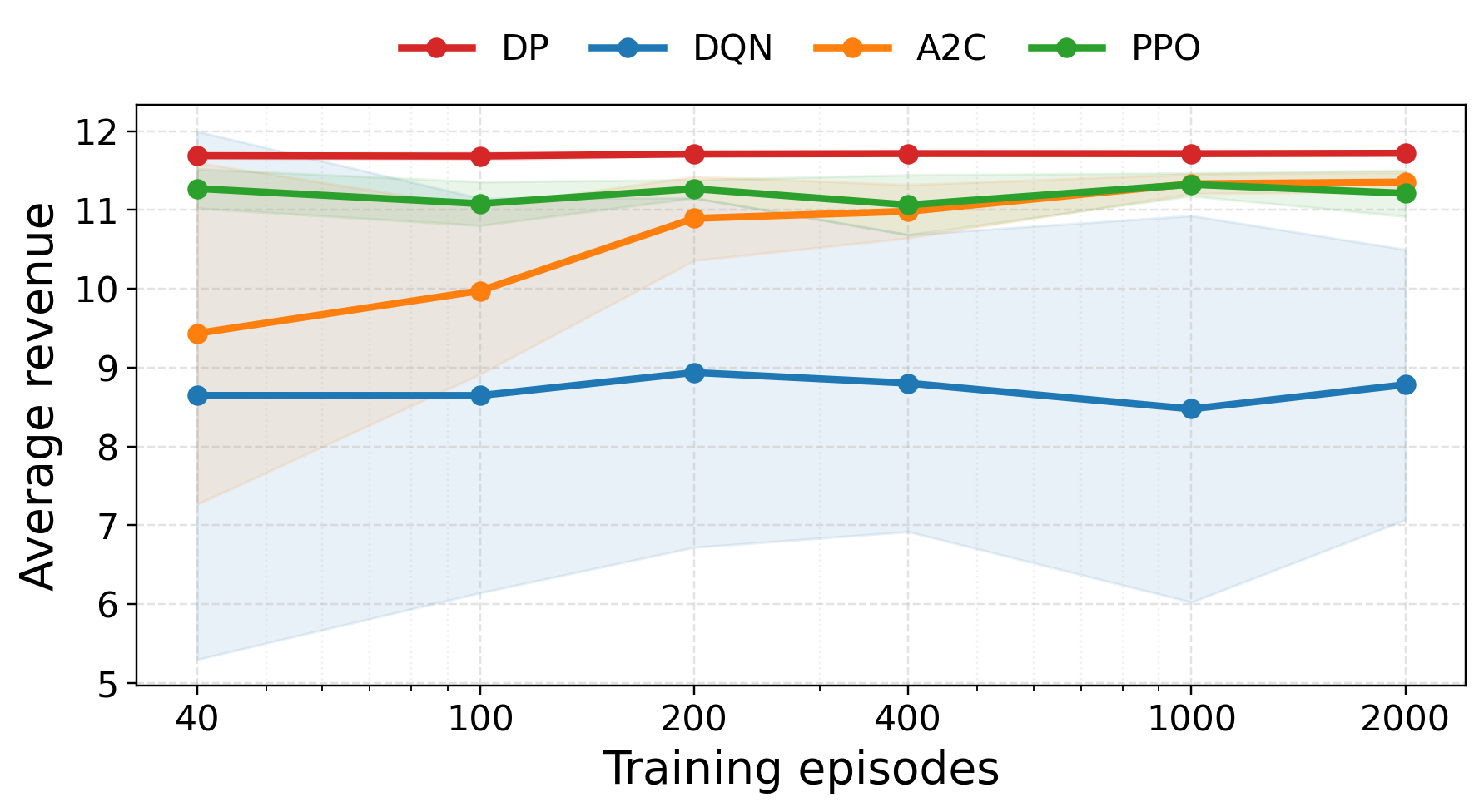}
    \caption{Environment 5: single typology with non-linear demand. Mean revenue and one standard deviation band over $10$ runs.}
    \label{fig:5_std}
\end{figure}

Figure~\ref{fig:5_std} shows the average revenue achieved by DQN, A2C, PPO, and Fitted DP as a function of the number of training episodes. The shaded regions correspond to one standard deviation over multiple runs.

The overall pattern is consistent with the main findings of the paper. Fitted DP achieves the highest and most stable performance across all training budgets. Its revenue remains almost constant, which reflects the fact that the Fitted demand model is directly optimized and does not rely on sampling-based policy improvement.

Among the RL methods, A2C shows the clearest learning trend. For small training budgets, its performance is noticeably below the DP benchmark, but it improves steadily as the number of training episodes increases and becomes competitive with PPO at larger budgets. PPO demonstrates strong and stable performance already at low training budgets, with relatively small variance throughout the experiment. DQN again underperforms the other methods and exhibits the largest variability.

Thus, the experiment with non-linear demand does not change the qualitative conclusions of the paper. Fitted DP remains the strongest method in terms of average revenue, PPO remains the most stable RL baseline, and A2C continues to benefit substantially from larger training budgets. This suggests that the main comparative conclusions are not limited to environments with linear demand specifications and extend to more complex non-linear settings as well.

\bibliographystyle{plain}
\bibliography{references}

\Addresses

\end{document}